\documentstyle[12pt,aasms4]{article}
\lefthead{Feuchtgruber et al.}
\righthead{Detection of interstellar ${\rm CH_3}$}
\slugcomment{Submitted to \apj\ Lett. [March 2000]}
\begin{document}

   \title{Detection of interstellar ${\rm CH_3}$
        \footnote
        {Based on observations made with ISO, a project of ESA with 
        participation of ISAS and NASA, and the SWS, a joint project of SRON
        and MPE with contributions from KU Leuven, Steward Observatory and 
        Phillips Laboratory.} 
         }

   \author{H.\ Feuchtgruber}
   \affil{Max-Planck-Institut f\"ur extraterrestrische Physik, Postfach 1603,
      D--85740 Garching, Germany}
   \authoremail{fgb@mpe.mpg.de}
   \author{F.\ P.\ Helmich}
   \affil{SRON, P.O. Box 800, 9700 AV Groningen, The Netherlands} 
   \author{E.\ F.\ van\ Dishoeck}
   \affil{Leiden Observatory, P.O. Box 9513, NL-2300 RA Leiden, 
          The Netherlands}       
   \author{C.\ M.\ Wright}
   \affil{School of Physics, University College, Australian Defence Force
      Academy, UNSW, Canberra ACT 2600, Australia}

   \begin{abstract} 
Observations with the Short Wavelength
Spectrometer (SWS) onboard the {\it Infrared Space Observatory} (ISO)
have led to the first detection of the methyl radical ${\rm CH_3}$ in
the interstellar medium.  The $\nu_2$ $Q-$branch at 16.5 $\mu$m and
the $R$(0) line at 16.0 $\mu$m have been unambiguously detected toward
the Galactic center SgrA$^*$. The analysis of the measured bands gives
a column density of (8.0$\pm$2.4)$\times10^{14}$ cm$^{-2}$ and an
excitation temperature of $(17\pm 2)$ K. Gaseous ${\rm CO}$ at 
a similarly low 
excitation
temperature and ${\rm C_2H_2}$ are detected for the same line of
sight. Using constraints on the ${\rm H_2}$ column density obtained from
${\rm C^{18}O}$ and visual extinction, the
inferred ${\rm CH_3}$ abundance is $(1.3{{+2.2}\atop{-0.7}}) \times 10^{-8}$. 
The chemically related ${\rm CH_4}$ molecule is not detected, but the pure
rotational lines of ${\rm CH}$ are seen with the Long Wavelength Spectrometer
(LWS).  The absolute abundances and the ${\rm CH_3/CH_4}$ and ${\rm CH_3/CH}$
ratios are inconsistent with published pure gas-phase models of dense
clouds. The data require
a mix of diffuse and translucent clouds with different densities and 
extinctions, and/or the development of translucent models in which gas-grain 
chemistry, freeze-out and reactions of ${\rm H}$ with polycyclic aromatic 
hydrocarbons and solid aliphatic material are included.

\end{abstract}


%

\section{Introduction}

The methyl radical ${\rm CH_3}$ is an important intermediate product
in the basic ion-molecule gas-phase chemistry networks in the
interstellar medium driven by cosmic-ray ionization (Herbst \&
Klemperer 1973). Together with ${\rm CH}$ and ${\rm CH_2}$, it is
produced by a series of reactions starting with C + H$_3^+$ $\to$
CH$^+$ + H or the radiative association of C$^+$ + H$_2$ $\to$
CH$_2^+$ + h$\nu$, followed by hydrogen abstraction reactions and
dissociative recombination.  Alternatively, it can be produced by
photodissociation of methane (CH$_4$).
Subsequent reactions of C$^+$ with CH$_3$ form one of the most
important steps in the formation of more complex hydrocarbons.

The ion ${\rm H_3^+}$ which initiates the chemistry in the cold gas
has been detected only recently toward the Galactic center by Geballe
et al.\ (1999) at a surprisingly high abundance.  This unique line of
sight turns out to be an extremely valuable environment to study
abundances in the cold low density interstellar medium, since even minor 
species like ${\rm CH_3}$ may be detectable due to its long path.
   
Although ${\rm CH_3}$ is a simple species, it is difficult to obtain
accurate laboratory measurements of its molecular parameters since, as
a radical, it recombines very fast with other particles in a
gas.  Herzberg (1961) and Herzberg \& Shoosmith (1956) were the first
to determine that the molecule is planar, but definite proof came only
from measurements of the out-of-plane bending mode $\nu_2$ at 16
$\mu$m by Yamada et al.\ (1981).
 
Observations of the 16--16.5 $\mu$m wavelength range are strongly
hampered from the ground due to the Earth's atmosphere. The first
detections of CH$_3$ in space have become possible only using the {\it
Infrared Space Observatory} (ISO) (Kessler et al.\ 1996).  B\'ezard et
al.\ (1998, 1999) have recently detected CH$_3$ in the atmospheres of
Saturn and Neptune respectively, but no previous searches for the
molecule in interstellar space have been reported.

\section{${\rm CH_3}$ spectroscopy}

Since the CH$_3$ radical is planar and symmetric, it does not have
electric dipole allowed rotational lines which could be detected in
the (sub-)millimeter wavelength range. The planar nature also implies
that the symmetric stretch $\nu_1$ is infrared inactive and the
asymmetric stretch $\nu_3$ at 3.16 $\mu$m relatively weak. Indeed, the
transition dipole moment of the $\nu_3$ band is found to be a factor of
three weaker than the out-of-plane bending mode $\nu_2$ (Triggs et
al.\ 1992, Amano et al.\ 1992).

To calculate the $\nu_2$ spectrum, the term energies were taken from
Yamada et al.\ (1981). The nuclear spin of the H-atoms can couple
either to a quartet or doublet state, with nuclear spin statistical
weights of 4 and 2, respectively. Because CH$_3$ follows Fermi-Dirac
statistics, the $K = 3,6,9,\ldots$ levels are quartet states, and the
other K-values doublet states. The strongest $Q$-branch has $N = K$
and is located at 16.5~$\mu$m; the strongest other feature is the
$^{Q}R(0)$ line at 16.0 $\mu$m thanks to its favorable H\"onl-London
factor. The band strength of $(2.5\pm
0.8)\times 10^{-17}$ cm$^{-1}$ (molecule cm$^{-2}$)$^{-1}$ was taken
from Wormhoudt \& McCurdy (1989). The calculation of the spectrum was
performed as described in Helmich (1996).

The shape of the spectrum is very sensitive to the excitation
temperature (see Figure 7.14 of Helmich 1996). 
Besides the strong ${^QQ}$-branch and $^{Q}R(0)$
lines, many more features become visible  at excitation temperatures above
25~K, most notable the satellite ${^RQ}$-branch at 16.53~$\mu$m
and the $P$(2) line at 17.60 $\mu$m.

\section{Observations and Data Reduction}

Observations were carried out in the SWS grating mode AOT06 (de Graauw
et al.\ 1996) at a spectral resolving power $R= \lambda / \Delta
\lambda \approx 1500-2200$.  The spectral range covering the ${\rm
CH_3}$ $Q$-branch and the $P$(2) line has been measured on
1997 February 21 15:24:27-19:03:01 UT, whereas that covering the $R$(0) and
$R$(1) lines has been obtained on the same day at 19:03:45-20:55:41
UT.  The SWS aperture size was ${\rm 14''\times27''}$ and has been
centered on the position of Sgr A$^*$ RA $17^h 45^m 40^s.0$, 
Dec $-29^{\circ} 00' 28''.6$ (J2000 coordinates), with the long side of the
slit oriented within 1 degree of the north-south direction. Due to the rather
large aperture size, the Galactic center sources IRS 1, 2, 3 and 7 also fall
inside the beam (see, e.g., Geballe et al.\ 1989), whereas IRS 5 and 6 are
positioned just outside the slit.

Data were processed within the SWS interactive analysis system, based
on the standard ISO pipeline OLP V8.7 products.  The data reduction
adhered to the recommendations of Salama et al.\ (1997).  Raw data
were rebinned to $R=5000$, a value significantly larger than the
actual spectral resolving power of the SWS to avoid losing spectral
detail when convolving the observed data samples by the bin.

The absolute calibration of the SWS data has about $\pm20\%$
uncertainty on average longwards of $\sim$15 $\mu$m (Salama et al.\
1997). However, since our analysis is entirely based on spectra where
the continuum is divided out, the actual uncertainty in the results is
determined by the noise in the data rather than the actual calibration
uncertainty. The main limitation of our analysis originates from the
$\pm30\%$ uncertainty in the ${\rm CH_3}$ $\nu_2$ band strength
(Wormhoudt \& McCurdy 1989, Yamada \& Hirota 1983).

\section{CH$_3$ results}

As shown in Fig.~1, the $Q$-branch at 16.5 $\mu$m and
the $R$(0) line at 16.0 $\mu$m are clearly detected. This represents
the first unambiguous detection of ${\rm CH_3}$ in the interstellar
medium. The upper
limit on the $P$(2) line at 17.60 $\mu$m provides an important
constraint on the temperature. Due to a blend with the [Ne\,III]
15.555 $\mu$m atomic fine structure line at the SWS grating
resolution, no information from the $R$(1) line at 15.54 $\mu$m could
be obtained.

Both the $Q$-branch and the $R$(0) lines are shifted with respect to
their expected LSR wavelengths by about $-20$ km s$^{-1}$.
Although such a shift is close to the SWS
wavelength calibration accuracy (Valentijn et al.\ 1996), a $V_{\rm
LSR}=-30$ km s$^{-1}$ component of cold molecular gas has been
reported previously by several authors from observations at radio and
millimeter wavelengths with similar beam sizes (Serabyn et al.\ 1986
and Sutton et al.\ 1990, ${\rm CO}$; Pauls et al.\ 1996, 
${\rm H_2CO}$; Serabyn \& G\"usten 1986, ${\rm NH_3}$; Marr
et al.\ 1992, ${\rm HCO^{+}}$; G\"usten et al.\ 1987, ${\rm HCN}$;
Bolton et al.\ 1964, ${\rm OH}$). In all cases 
several velocity components at
$\sim -50, -30$ and 0 km s$^{-1}$ have been observed at much higher
spectral resolutions. The relative strengths of these three components
vary between the observed species with the 0 km s$^{-1}$ component
often the largest. At the SWS spectral resolution of $\sim 150$ km
s$^{-1}$ it is not possible to distinguish between these different
velocity components, but our observed shift is consistent with a mix
of them. The location of the absorbing gas can therefore not be
attributed to one particular feature, but is possibly spread along the
line of sight toward Sgr A$^*$ among spiral arms and molecular clouds.

Fits of synthetic spectra to the data, as described in \S 2 and matching 
the SWS resolution, were performed
for different excitation temperatures (10 to 50~K), different Doppler
parameters ($b$ between 1.5 and 30 km~s$^{-1}$) and column densities.
The best fit to the individual absorption depths
(Fig.~1\ a,b) as well as their ratio
(Fig.~1\ c) is given in Table~1. The absorption depths
are almost independent of Doppler parameter, and are mainly a function
of column density. The ratio of the depths is a strong function of the
excitation temperature. The inferred low excitation temperature of
$(17\pm 2)$~K from the 16.0/16.5 $\mu$m ratio of (0.8$\pm$0.15)
is consistent with the non-detection of the P(2) line. 
Because ${\rm CH_3}$ has no dipole moment, the populations of the lowest
rotational levels are controlled by collisions, so that the excitation
temperature is close to the kinetic temperature.

The H$_2$ column density along the line of sight has been constrained
by several sets of observations. First, the measured extinction of 31
mag (Rieke et al.\ 1989) implies $N$(H$_2$)$\approx 2\times 10^{22}$
cm$^{-2}$ using $N_H/A_V=1.9 \times 10^{21}$ cm$^{-2}$ mag$^{-1}$ and
assuming that at least half of the hydrogen is in molecular form.
Second, the detection of at
least one optically thin C$^{18}$O line (R(0) at 4.7716 $\mu$m) in our ISO-SWS 
observation together with the measured CO excitation 
temperature of 8-13 K implies $N$(C$^{18}$O)=$(2\pm 0.5)\times 10^{16}$ 
cm$^{-2}$. This is in good agreement with the analysis by 
Moneti \& Cernicharo (2000) on the same data. 
Using $^{16}$O/$^{18}$O= 300 (Wilson \& Rood 1994)
and CO/H$_2$=$10^{-4}$ implies 
$N$(H$_2$)$\approx6\times 10^{22}$ cm$^{-2}$. 
We adopt $N$(H$_2$)=$(6\pm 3)\times 10^{22}$ cm$^{-2}$, leading to a CH$_3$
abundance with respect to H$_2$ of $x$(CH$_3$)=$(1.3{{+2.2}\atop{-0.7}}) \times
10^{-8}$. Note that all abundances would be increased by a factor of 3
if the lower H$_2$ column density derived from the extinction is
used.

\section{Related species: CH$_4$, C$_2$H$_2$ and CH}

The availability of the full SWS scan allows searches for other
chemically related molecules. Specifically, the $\nu_2/\nu_4$ dyad of
CH$_4$ occurs around 7.7 $\mu$m and has been observed with the ISO-SWS
toward massive protostars by Boogert et al.\ (1998).  Toward the
Galactic center, however, gas-phase CH$_4$ is not detected.  Adopting
the same excitation temperature as found for ${\rm CH_3}$, an upper
limit for its abundance of $\le$1$\times10^{15}$ cm$^{-2}$ is found.
Solid ${\rm CH_4}$ is clearly detected by Chiar et al.\ (2000) toward Sgr
A$^*$ with a column density of ($3.0\pm 0.7$)$\times 10^{16}$
cm$^{-2}$. Thus, most of the CH$_4$ is in solid form, consistent with
the low temperature.

Detection of a blend of the pure rotational lines of ${\rm CH}$ at 149.09 and
149.39 $\mu$m towards the Galactic center has been reported by White
et al.\ (1999). We have re-analyzed the LWS observations carried out
on 1998 February 20 10:11:34-11:06:44 in the LWS01 grating mode (Clegg et
al.\ 1996) at a resolution of $\sim$ 1500 km s$^{-1}$ 
(Fig.~2\ a). The LWS data
reduction has been based on OLP V8.7 products and has been carried out
within ISAP (Sturm et al.\ 1997). Outliers have been removed by
iterative sigma clipping and the different scans have been flatfielded
to their mean value by applying a second order polynomial offset to
each individual scan. The fringing in the LWS data, present in all
extended source observations has been removed by the dedicated module
within ISAP.
The inferred equivalent width for the unresolved doublet is $(85\pm
5)$ km s$^{-1}$. Using the formulae from Stacey et al.\ (1987) and
assuming $T_{\rm ex}$=17~K, this leads to $N$(CH)$\approx (1.1 \pm
0.1) \times 10^{15}$ cm$^{-2}$.

Finally, the $\nu_5$ band of gas-phase C$_2$H$_2$ at 13.7 $\mu$m is
clearly detected (Fig.~2\ b). Following the analysis 
of Lahuis \& van Dishoeck (2000), we 
find $N$(C$_2$H$_2$)=($5.5 \pm 0.8)\times 10^{14}$
cm$^{-2}$ assuming $T_{\rm ex}=17$~K.  Table~1
summarizes the results obtained from ISO observations. Note that the relative 
abundances of the molecules have much smaller error bars than the absolute 
values since the uncertainty in the ${\rm H_2}$ column density cancels out.

\section{Chemistry}

The absolute and relative abundances of the observed molecules have
been compared with a wide variety of models, including time- and
depth-dependent models.  None of the published pure gas-phase
dense cloud models can reproduce the observations of all species 
(e.g., Millar et al.\ 1997, Lee et al.\ 1996, see
Table~2). In general, the model CH$_3$ abundances are too low and the
CH$_4$ abundances too high.  Also, the model abundance of C$_2$H$_2$
is significantly smaller than that of CH$_4$, in contrast with
observations.  The only models which come close to matching the
absolute and relative abundances of CH$_3$ and CH$_3$/CH and
CH$_3$/CH$_4$ are low-density translucent cloud models with
$n$(H$_2$)$\approx 10^3$ cm$^{-3}$ and $A_V \approx $few mag, so that
photodissociation of CH$_4$ to CH$_3$ and CH$_2$ is still
efficient. Table~2 lists recent model calculations by Terzieva \&
Herbst (priv.\ communication) and results based on the models by
van Dishoeck \& Black (1986) and
Jansen et al.\ (1995) using updated branching ratios for the
dissociative recombination of the hydrocarbon ions (Andersen et al.\
2000).  Low metal abundances are favored, to prevent destruction of
the hydrocarbons by oxygen and by
sulfur atoms and ions.  Note, however, that even
though the abundances at $A_V\approx 3$ may match the data
within a factor of a few, the CH$_3$
column density in such models is
only $1\times 10^{13}$ cm$^{-2}$, nearly two
orders of magnitude below observations. At the same time, the CH
column density of $1.4\times 10^{14}$ cm$^{-2}$ is a factor of 10 below
observations.
Because of the strong depth dependence of the CH, CH$_3$ and CH$_4$ abundances,
it is not possible to reproduce the column density ratios with these same
models.
The large observed ${\rm H_3^+}$ column density suggests that there are several 
clouds along the line of sight. Some combination of low-density diffuse 
clouds to produce the CH and denser clouds to account for the solid CH$_4$
may explain those data, but the mix would have to be tailored very 
specifically to simultaneously approach the large column densities of CH$_3$ 
and C$_2$H$_2$.

An alternative suggestion is to invoke turbulent chemistry, in which a high 
${\rm CH^+}$ abundance leads to enhancements of other hydrocarbons by 1-2 
orders of magnitude (e.g. Hogerheijde et al.\ 1995, Joulain et al.\ 1998).
However, the relative ratios of CH and CH$_3$ are unlikely to change 
in such models.

Given the detection of solid CH$_4$ and the low inferred temperatures,
it is plausible that gas-solid interactions and grain-surface chemistry
also play a role in producing the hydrocarbons. In this respect, the
situation for CH$_3$ may be similar to that for NH in diffuse clouds
(Mann \& Williams 1984, van Dishoeck 1998). Conversion of atomic
carbon to small hydrides on grain surfaces  
may be significant, but no model results
exist yet for these conditions. Such models should also explain the
C$_2$H$_2$ abundances and lack of complete C$_2$H$_2$ freeze-out.
Alternatively, reactions of atomic H with PAHs and solid aliphatic 
hydrocarbon material, known to be present toward Sgr A$^*$ from the 
3.4 $\mu$m absorption feature, may lead to CH$_3$.
Shock chemistry is not likely to be important for this line of sight
because of the low temperatures. 

Future high spectral resolution observations
of CH$_3$ toward Sgr A$^*$ to constrain the velocity structure, as well 
as observations of CH$_3$ and other molecules in different types of diffuse
clouds are needed to constrain the basic hydrocarbon chemistry.

\acknowledgements
The authors are grateful to the SWS instrument teams, to 
E.\ Herbst, R.\ Terzieva and D.J.\ Jansen for updated model results 
on CH$_3$ and to W. Duley for inspiring discussions. This work was 
supported by DARA grants no 50 QI9402 3 and 50 QI 8610 8 and by NWO 
grant 614.41.003. CMW acknowledges receipt of an ARC Australian 
Postdoctoral Fellowship.

\newpage

\begin{table} 
   \caption{Summary of results}
   \label{Table 1}
   \begin{tabular}{lcclc}
   \hline\\[-2.0ex]
    Molecule & Col. Dens.& T$_{ex}$ & Obs. Mode$^a$ \\[0.4ex] 
             &   [cm$^{-2}$]& [K]    \\[0.4ex]\hline
    & &   \\[-1.7ex]     
    ${\rm CH}$   & (1.1$\pm$0.1)$\times10^{15}$& 17$^b$  & LWS01   \\[0.3ex]
    ${\rm CH_3}$ & (8.0$\pm$2.4)$\times10^{14}$& (17$\pm$2) & SWS06  \\[0.3ex]
    ${\rm CH_4}$ & $\le$1$\times10^{15}$& 17$^b$       & SWS06  \\[0.3ex]
    C$_2$H$_2$ & (5.5$\pm$0.8)$\times 10^{14}$ & 17$^b$ & SWS06\\[0.3ex]
    C$^{18}$O  & (2.0$\pm$0.5)$\times 10^{16}$& 8-13 & SWS06\\[0.6ex]
\hline
\end{tabular}
\phantom{}\\[1.0ex]
\smallskip
$^a$ For details of the observing modes see de 
    Graauw et al.\ (1996) and Clegg et al.\ (1996) \\
$^b$ Assumed  excitation temperature
\end{table}

\newpage

\begin{table*} 
   \caption{Comparison of observed abundances with models$^a$}
   \label{models}
   \begin{tabular}{lrrrr}
   \hline\\[-2.0ex]
    Model & CH & CH$_3$ & CH$_4$ & C$_2$H$_2$ \\[0.4ex]\hline
    & &   \\[-1.7ex]     
 Obs.   &18${{+22}\atop{-7}}$(-9)&13${{+22}\atop{-7}}$(-9)&$<$17(-9)
        &9${{+21}\atop{-4}}$(-9)\\[0.3ex]
 LBH96$^b$ & 0.26(-9) & 0.20(-9) & 180(-9) & 15(-9)\\[0.3ex]
 TH00$^c$ & 44(-9) & 8.6(-9) & 5.6(-9) & 9.2(-9)\\[0.3ex]
 VDB00$^d$ & 3.2(-9) &6.4(-9) & 18(-9) & 2.0(-9)\\[0.6ex]
\hline
\end{tabular}
\phantom{}\\[1.0ex]
\smallskip
$^a$ All abundances with respect to H$_2$ \\
$^b$ Lee et al.\ (1996) new standard model with $n_{\rm H}=10^4$ cm$^{-3}$
and $T$=10 K at steady state with low metal abundances \\
$^c$ Terzieva \& Herbst (priv.\ comm.) translucent cloud model with
$n_{\rm H}=2\times 10^3$ cm$^{-3}$, $T$=10 K and $A_V$=3 mag 
at steady state with low
metal abundances \\
$^d$ Updated models of van Dishoeck \& Black
(1986) and Jansen et al.\ (1995) for $n_{\rm H}=2\times 10^3$
cm$^{-3}$ and $A_V$=3 mag at steady state with low metal abundances \\
\end{table*}

\newpage

\figcaption{(a),(b): Data and synthetic spectra of the ${\rm CH_3}$
         observations; (c): Determination of the CH$_3$ excitation temperature
         $T_{\rm ex}$ from the integrated absorption ratio of the 16.0 $\mu$m
         $R$(0) line and 16.5 $\mu$m $Q$-branch.}

\figcaption{ISO spectra of Sgr A$^*$:
(a) LWS spectrum of the CH pure rotational line doublet; (b) SWS spectrum
    around the C$_2$H$_2$ $\nu_5$ Q-branch and the synthetic spectrum.}

\end{document}